\documentclass[prl,aps,twocolumn,showpacs]{revtex4}
 \usepackage{graphicx}
\newcommand{\gsim}{\gtrsim}
\newcommand{\lsim}{\lesssim}
\newcommand{\arccosh}{{\rm arccosh}}

\begin{document}
 
 \title{Density profile of a harmonically trapped ideal Fermi gas in arbitrary dimension}
 \author{Erich J. Mueller}
\affiliation{Laboratory of Atomic and Solid State Physics, Cornell University, Ithaca, New York 14853}
\date{May 21, 2004}

\begin{abstract}
Closed form analytic expressions are derived for the density profile of a harmonically trapped noninteracting Fermi gas in $d$ dimensions.  Shell structure effects are included to leading order in $1/N$, where $N$ is the number of particles.  These corrections to the local density approximation scale as $\delta n/n\sim N^{-\alpha}$, where $\alpha=(1+1/d)/2$.  
\end{abstract}
 \maketitle
 Experiments on quantum degenerate Fermi atoms have motivated a series of theoretical studies of the basic properties of zero \cite{minguzzi1, march,brack,gleisberg,brack2,minguzzi2} and finite \cite{zyl1} temperature noninteracting fermions in inhomogeneous potentials.  Here we derive closed form analytic expressions for the ground state density of a harmonically trapped Fermi gas in $d$ dimensions.

 Theoretical studies of noninteracting particles are crucial for understanding experiments on nonresonant Fermi gases \cite{foot1}.  A typical length scale for interactions is $r_0\approx$nm, and a typical interparticle spacing is $n^{-1/3}\approx \mu$m$\gg r_0$.  Thus, even in multiple component gases, where $s$-wave collisions are allowed, the interaction energy per particle $E_{\rm int}\approx \hbar^2 r_0 n/m$ is small compared with the Fermi energy $E_f\approx \hbar^2 n^{2/3}/m$.  Therefore the interactions can be ignored for calculating gross features of the ground state structure.  This separation of energy scales is even more  dramatic in a spin-polarized gas, where $s$-wave collisions are forbidden, and therefore $p$-wave collisions dominate.  The cross-section for $p$-wave collisions is down by a factor of $(k_f r_0)^4$ relative to $s$-wave.  

Although our calculations are for arbitrary dimension, $d$, a particularly important case is $d=1$ where theoretical predictions about noninteracting fermions can be applied to hard-core bosons \cite{girardeau}.
 
Most experiments are performed in harmonic traps, so it is natural to consider the density profile of a gas of $N$ fermions in a potential $V(r)= m\omega^2 r^2/2$.  
 Since we ignore interactions, each spin component is independent, 
 and it suffices to consider the spinless (or spin-polarized) case.
 The simplest expression for this density profile comes from the local density (Thomas-Fermi) approximation, where the trap potential gives a spatially dependent chemical potential $\mu(r)=k_f^2(r)/2m=\mu_0-V(r)$, and the local density is derived from the relationship between density and chemical potential in a $d$-dimensional homogeneous system, $n=(k_f/2\sqrt{\pi})^{d}/\Gamma(d/2+1)$, resulting in a density profile,
 \begin{equation}\label{ld}
 n_{\rm tf}(r)= \frac{(R_{\rm tf}^2-r^2)^{d/2}}{2^d\pi^{d/2}\Gamma(d/2+1)\ell^d}
 \end{equation}
 where $R_{\rm tf}^2/\ell^2=2 (d!N)^{1/d}$ is the radius of the cloud,and $\ell^2=\hbar^2/m\omega^2$ is the oscillator length.
Although this is an excellent approximation to the true ground-state density profile, it misses ``shell effects" resulting from the statistical correlations in an ideal Fermi gas.  These shell effects give rise to corregations on top of this smooth profile which are analogous to the Friedel oscillations which occur in the density of a uniform Fermi gas near an impurity \cite{friedel}.  As we show below, these corregations scale as $\delta n/n\sim N^{-\alpha}$
where $\alpha=(1+1/d)/2$.  These deviations are therefore only significant for small numbers of particles.  Despite their small size, previous estimates \cite{minguzzi1,minguzzi3} suggest that these deviations are experimentally observable.

Several authors have quantified these density fluctuations.
%deviations from the local density expression (\ref{ld}).  
In one dimension,   Husimi \cite{husimi} 
%used the Christoffel-Darboux formula to 
derived a simple closed-form expression for the {\em exact} density profile in terms of Hermite polynomials.
Further discussion and applications 
of this approach 
%and relationships to other approximations 
can be found in \cite{damski}.
%Damski et al. \cite{damski} and Kim and Zubarev \cite{kim} recently made use of this expression.
Using a completely independent formalism,
Vignolo et al. \cite{minguzzi1} recently derived an efficient numerical technique for calculating this profile.  Vignolo and Minguzzi later generalized their numerical approach to higher dimension \cite{minguzzi2}.  Also in higher dimensions, Brack and Zyl \cite{brack} derived an analytic expression for the exact density profile in terms of a sum over Laguerre polynomials.   Brute force numerical calculations of three dimensional shell effects can be found in \cite{brute}, including some discussion of corrections due to interactions.
%A partial list of finite temperature results in various geometries can be found in references \cite{zyl1,zyl2,wang}.

Unlike these prior works, we will not attempt to derive an exact expression for the density profile, rather we will produce an asymptotic series in powers of $1/N$.  Consequently we are able to derive simple analytic formulae.  Near the center of the trap, our results agree with those derived by Gleisberg et al \cite{gleisberg} and Brack and Murthy \cite{brack2}.  Our expressions match the exact density profile throughout the cloud, even for surprisingly small numbers of particles: $N\sim 1$.

\begin{figure}
\begin{centering}
\includegraphics[width=0.9\columnwidth]{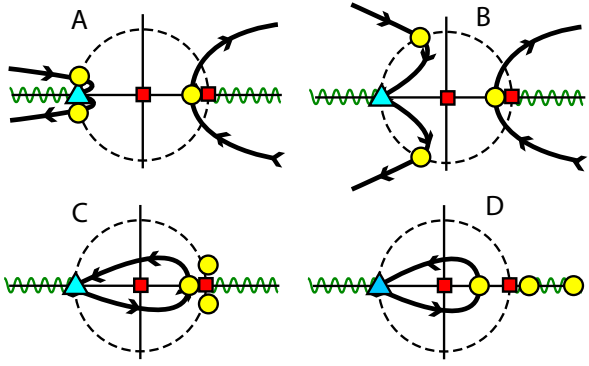}
\end{centering}
\caption{(color online) Contours of steepest descents in the complex $t$ plane for regions A,B,C,D (see text).  Poles (at $t=0,1$) are represented by red squares, saddle points by yellow discs,  essential singularities (at $t=-1$) by blue triangles, and branch cuts (extending to $\pm\infty$ from $t=\pm 1$) by green wiggly lines.  The dashed circle corresponds to $|t|=1$.}\label{paths}
\end{figure}

\noindent{\bf Density Generating Function:}
The single particle states of a $d$ dimensional harmonic oscillator are described by the quantum numbers $\{n_1,\ldots,n_d\}$, corresponding to a real-space wavefunction of the form $\psi(x_1,\ldots,x_d)=\phi_{n_1}(x_1)\cdots\phi_{n_d}(x_n)$, where $\phi_j(x)= H_j(y)e^{-y^2/2}/(2^j j! \ell \sqrt{\pi})^{1/2}$ is the eigenfunction of the one-dimensional harmonic oscillator, the dimensionless length is $y=x/\ell$, and $H_j(y)$ is the $j$'th Hermite polynomial.  This single-particle state has
 energy $E=\hbar \omega \sum_j (n_j+1/2)$. The ground state
 of $N$ fermions in a harmonic trap is only unique if an integral number of shells are filled, requiring 
\begin{equation}
N= {{M+d}\choose{d}}\approx \frac{M^d}{d!}\left(1+\frac{d (d+1)}{2 M}+{\cal O}(M^{-2})\right),
\end{equation}
 where $M=0,1,2,\ldots$ indexes the shell.  The Thomas-Fermi radius is then $R_{\rm tf}^2/\ell^2=
 2M+(1+d)+{\cal O}(1/M)$.   The exact density will be denoted $\rho_M^{(d)}({\bf r})=\sum_{m=1}^M\sigma_m^{(d)}({\bf r})$, where $\sigma_m^{(d)}({\bf r})=\sum_{n_1+..+n_d=m}|\phi_{n_1}(x_1)\cdots\phi_{n_d}(x_d)|^2$ is the density contribution from the $m$'th shell.  In the remainder of this paper, we scale all lengths by $\ell$.
 
% The superscript $(d)$ in $\rho$ and $\sigma$ will usually be omited.
  
We will extract the large $M$ structure of the density from the generating function, $G({\bf r},t)=\sum_M \rho_M^{(d)}({\bf r}) t^M$.  We construct $G$ by noting that the Hermite polynomials obey 
\begin{equation}
F(x,z)=\sum_j\frac{z^j}{j!} H_j(x) = \exp(-z^2+2 z x).
\end{equation}
To construct expressions involving the density contribution from each orbital, $|\phi_j(x)|^2$, we need a generating function for $H_j(x)^2$.  We therefore introduce
\begin{eqnarray}\nonumber
\bar W(x,s)&\equiv&\!\!\int\!\!du\!\!\int\!\!dv\, e^{-(u^2+v^2)/s^2} F(x,u+i v) F(x,u-i v)\\
&=&\pi \sum_j \frac{H_j(x)^2 s^{2 j+2}}{j!}\\\nonumber
&=&\frac{\pi s^2}{\sqrt{1-4 s^2}} e^{4 s^2 x^2/(1+2 s^2)}.
\end{eqnarray}
A simple rescaling then gives
\begin{equation}
W(x,t)=\sum_j |\phi_j(x)|^2 t_j=\frac{2}{\pi^{3/2} t} e^{-x^2} \bar W(x,\sqrt{t/2}).
\end{equation}
The generating function for the contribution to the density in $d$ dimensions from the $M$'th shell is the product
\begin{equation}
\bar G({\bf r},t)=\sum_m \sigma_m^{d}({\bf r}) t^m =
\prod_{j=1}^d W(x_j,t).
\end{equation}
The density for $M$ filled shells then obey
\begin{eqnarray}\label{fin}
G({\bf r},t)&=&\sum_M \rho_M^{(d)}({\bf r}) t^M
=\frac{\bar G({\bf r},t)}{(1-t)}\\\nonumber
&=&\frac{1}{\pi^{d/2} (1-t)^{1+d/2}(1+t)^{d/2}}
\exp\left(-\frac{1-t}{1+t} r^2\right),
\end{eqnarray}
which shows (as would be expected from symmetry) that the density is only a function 
of the distance $r=|{\bf r}|$.

\noindent {\bf Asymptotic Expansion:}
The relationship (\ref{fin}) is inverted by performing a contour integral around the origin in the complex $t$ plane,
\begin{equation}\label{densexp}
 \rho_M^{(d)}(r)=\frac{1}{2\pi i} \oint\!\frac{dt}{t^{M+1}} G(r,t).
\end{equation}
Deforming the contour to follow the path of steepest descents, the large $M$ asymptotics are calculated from the portion of the integral which passes over saddle points.  The path of steepest descents is illustrated in figure~\ref{paths} for four different regimes:  $A: r^2/R_{\rm tf}^2\lsim R_{\rm tf}^{-4},$ $B: R_{\rm tf}^{-4}\lsim r^2/R_{\rm tf}^2\lsim1-R_{\rm tf}^{-4/3},$ 
$C: |r^2/R_{\rm tf}^2-1|\lsim R_{\rm tf}^{-4/3},$
and $D: r^2/R_{\rm tf}^2-1\gsim R_{\rm tf}^{-4/3}.$  These regimes are distinguished by the number and locations of the saddle points which are found on the steepest descents contour.
%(* Verify *)
%In counting powers of $M$, we take $x^2$ to be of order $M$.

In all cases the integral is dominated by a single saddle point on the real axis at $t=t_0$ with $0<t_0<1$.  For $r/R_{\rm \rm tf}<1-{\cal O}(R_{\rm \rm tf}^{4/3})$, this saddle point approaches the pole at $t=1$ as $t_0=1+{\cal O}(1/M)$, and one 
cannot use the standard saddle-point formulae to evaluate its contribution.  However,  
straightforward analysis shows that for a sufficiently well behaved \cite{behavior} function $f(t)$, one can approximate % such that the $n$'th derivative, $f^{(n)}(1)$, is of order $1$, and that $f(t)\to0$ sufficiently rapidly as $t\to\pm i \infty$, 
\begin{eqnarray}\label{theorem}
\frac{1}{2\pi i}\int\!\!dt\, \frac{e^{-M f(t)}}{(1-t)^\beta}=e^{-M f(1)}\frac{(Mf'(1))^{\beta-1}}{\Gamma(\beta)},
%&&\quad -\pi^{-1}(N f'(1))^{\beta-1}e^{-M f(0)}\sin(\pi\beta)\Gamma(1-\beta),
\end{eqnarray}
plus terms of relative order $1/M$.
The integral is taken on a contour from $t=-i\infty$ to $t=+i \infty$ with $\rm{Re}(t)<1$.  Near $t=1$, $G(r,t)/t^{M-1}$ has the same form as the integrand in (\ref{theorem}), with $M f(1)=(d/2)\log(2\pi)$ and $M f'(1)=(R_{\rm tf}^2-r^2)/2+{\cal O}(M^{0})$, where $R_{\rm \rm tf}^2= 2M+1+d$.
Thus the leading order behavior coincides with the Thomas-Fermi result in equation (\ref{ld}).

%
%invalidating the standard formulae for saddle point integration.
%
%saddle point approximation
%
%Setting the logarithmic derivative of $G(r,t)$ equal to zero, there are three saddle points at:
%\begin{eqnarray}
%t_0&=&1+{\cal O}(1/M)\\
%t_{\pm}&=&\frac{x^2}{M}-1\pm \frac{x}{M}\sqrt{x^2-2M}+{\cal O}(1/M).
%\end{eqnarray}
%As is apparent from evaluating $G$ at these points, the leading order structure comes from the integral near $t_0$.  Since $t_0$ approaches the pole at $t=1$ as $M\to\infty$, one 

%Examining the corrections to (\ref{theorem}), one finds that all contributions from the saddle point $t_0$ are smooth functions of $r^2$.  Density oscillations come solely from the contributions near the other two saddle points at $t=t_{\pm}$.  We must separately consider the four regions illustrated in figure~1:
The other saddle points are found by setting the logarithmic derivative of the integrand in (\ref{densexp}) equal to zero.  The resulting cubic equations yields two additional saddle points, denoted $t_\pm$.  In region A, $t_{\pm}\to-1$; in B, $t_{\pm}$ are isolated saddle points with $|t_\pm|\approx 1$; in C all three saddle points converge at $t=1$, and the contour of steepest descents no longer passes through $t_\pm$; and in D, all three saddle points lie on the real axis, and $t_0$ moves away from $t=1$, becoming an isolated saddle point.
%
%The other saddle points are at different locations depending upon the value of $x$.
%$A: x^2/R_{\rm tf}^2\lsim 1/\sqrt{M},$ where $t_{\pm}$ approaches the essential singularity at $t=-1$;
%$B: 1/\sqrt{M}<x^2/R_{\rm tf}^2<1-1/\sqrt{M},$ where $t_{\pm}$ are isolated saddle points; and 
%$C: x^2/2n\gsim 1-1/\sqrt{n},$ where the three saddle points converge at $t=1$ and the contour only passes through $t_0$; and $D: x^2/R_{\rm tf}^2\gsim 1+1/{R_{\rm tf}^2}$ where $t_0$ moves away from $t=1$, and becomes an isolated saddle point.
 In the last two regions we must explicitly consider the corrections to (\ref{theorem}).  Note that most of the cloud lies within region B and we therefore first consider the Friedel oscillations in this regime.  We then move on to regions $A$, $D$, and $C$.  This order is chosen so that we can sequentially use previous results.

%There are three regimes in which we need to calculate these corrections: (*Verify power of n*) $A: 1/\sqrt{M}<x^2/2M<1-1/\sqrt{M},$ $B: x^2/2M\lsim 1/\sqrt{M},$ and $C: x^2/2n\gsim 1-1/\sqrt{n}.$   The fluctuations in region $A$ can be evaluated by standard saddle-point techniques, and represents the "bulk" behavior away from the center (region B) and edges (region C) of the trap.  Near the center and edges of the trap,  the poles $t_{\pm}$ approach each other, requiring more sophisticated analysis.

\begin{figure}
\includegraphics[width=\columnwidth]{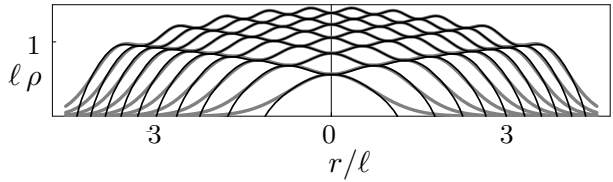}
\caption{Density profile of one-dimensional noninteracting fermions in a harmonic trap with frequency $\omega$.  Profiles with N=1 to N=11 particles are shown.  Density $\rho$ and position $r$ are measured in units of the trap length $\ell=\sqrt{\hbar/m\omega}$.  Thick gray lines: exact result; thin black lines: Eq. (\ref{res1}).  The exact and approximate results lie on top of one another except near the edge of the cloud.}\label{oned}
\end{figure}

\noindent {\bf Region B.}
Integrals passing over isolated saddle points can be evaluated through the standard formula,
%As long as the poles $t_\pm$ are sufficiently far apart, we can use the standard%
% standard saddle point formula
\begin{equation}\label{sadint}
\int_{\bar s} \!\!ds\,e^{M f(s)} =\sqrt{\frac{2\pi}{-M f''(\bar s)}}e^{M f(\bar s)}(1+{\cal O}(1/M^{1/2})),
\end{equation}
where $\bar s$ is the saddle point, characterized by $f'(\bar s)=0$, and the integral on the left represents the contribution to the integral along the path of steepest descents near the $\bar s$.  To simplify the algebra we change variables, letting $t=e^{2 i s}$.  Setting the logarithmic derivative of $G$ equal to zero, one finds that the saddle points $t_{\pm}$ correspond to $s_\pm=\pm\arccos(r/R_{\rm tf})+{\cal O}(1/M)$.
In this new variable, the integral in (\ref{densexp}) maps onto (\ref{sadint}) with
$M f(s)=i r^2 \tan s- i s R_{\rm tf}^2  - (d/2+1) \log(\pi) -(d/2)\log(2 \cos(s))-(1+d/2)\log(2 i \sin(s))$.  Adding the contributions of these two saddle points, one arrives at out central result: away from the center and edges of the trap, the leading order form of the density oscillations is
\begin{eqnarray}\label{res1}
\delta \rho_B &=& A_B\,
\bar r^{\frac{1-d}{2}} \left(1-\bar r^2\right)^{\frac{-(3+d)}{4}}
\cos(R_{\rm \rm tf}^2 \phi-\phi_0)\,\,\,\,\,\\
\phi&=&
-\bar r \sqrt{1-\bar r^2}+\arccos\bar r\\
A_B&=& 2^{-d}\pi^{-(1+d)/2} R_{\rm \rm tf}^{-1}\\
\phi_0&=&\pi (3+d)/4\\
\bar r &=& r/R_{\rm \rm tf} = r/(\sqrt{2 M})(1+{\cal{O}}(1/M)).
\end{eqnarray}
By examining Eq. (\ref{res1}), one sees that $\delta\rho$ scales as $1/\sqrt{M}$ for large $M$, while the Thomas-Fermi density scales as $M^{d/2}$.
The accuracy of (\ref{res1}) is illustrated by the examples shown in figure~\ref{oned}.

\noindent {\bf Region A.}  When $r\to0$ the poles $t_{\pm}$ approach one another at $t=-1$, and no longer contribute as isolated saddle points.  
The width of each saddle point scales as $(\Delta s)^2\sim R_{\rm tf}^3/r$ as $r\to 0$, while the distance between the saddle points ($s_+-s_-\sim\arccos(r/R_{\rm tf})$) approaches zero as $r/R_{\rm \rm tf}$.  Therefore the isolated saddle point approximation breaks down when $r^2<R_{\rm tf}^{-2}$.

To analyze this limit, we expand $G$ for small $r$ and $u=t+1$.  The leading order behavior for $\delta\rho$ is
\begin{equation}
\delta\rho_A=\frac{(-1)^{M+1}}{2^{1+d/2}}
\frac{e^{r^2}}{2\pi i}\int \frac{du}{u^{d/2}}\exp\left(-\frac{2 r^2}{u}+w u
\right),
\end{equation}
with $w=M+3/2+d/4\approx R_{\rm tf}^2/2$.
By deforming the contour, this integral may written in terms of a Bessel function,
$J_\mu(z)=(2\pi i)^{-1}\int du\, u^{-\mu-1}\exp(z(u-1/u)/2)$, to yield,
\begin{eqnarray}\label{rhoa}
\delta\rho_A&=&
\frac{(-1)^M}{2^{d}\pi^{d/2}}\bar r^{1-d/2} J_{d/2-1}(2\bar r R_{\rm \rm tf}^2).
\end{eqnarray}
This result was previously found in $d$ dimensions by Brack and Murthy \cite{brack}, and in 1 dimension by Gleisberg et al. \cite{gleisberg}.
By considering the long distance asymptotic behavior of the Bessel function $J_\nu(z)\sim\sqrt{2/\pi z}\cos(z-\pi\nu/2-\pi/4)(1+{\cal O}(z^{-2}))$, one sees that for $R_{\rm tf}^{-2} \ll\bar r\ll R_{\rm tf}^{-2/3}$ the solutions $\delta\rho_A$ and $\delta\rho_B$ coincide.

\noindent {\bf Region D.}
 As one makes $r$ larger, the saddle points which contribute to $\delta\rho$ follow the circle $|t|=1$, and as $r\to R_{\rm \rm tf}$ they approach the pole at $t=1$.  
For $r>R_{\rm tf}+{\cal O}(R_{\rm tf}^{-1/3})$ all three saddle points move onto the real axis; two with $t>1$ and one with $0<t<1$.  The path of steepest descents only passes over the last of the saddle points.  The Thomas-Fermi approximation has no applicability here, so we calculate the full density, rather than the deviation from Thomas-Fermi.  Since the relevant saddle point is isolated,  we can use the standard formula from equation (\ref{sadint}).  Similar to our approach in region B, we change variables, setting $t=e^{2 w}$, so that the density is
\begin{eqnarray}
\rho_D (r) &=& \frac{1}{i 2^{1+d}\pi^{1+d/2} }\int dw\,e^{\bar\theta},
\end{eqnarray}
where $\bar\theta = -R_{\rm tf}^2 w + r^2\tanh w -(1+d/2)\log\sinh(-w)-(d/2)\log\cosh w$.  To leading order, the saddle point is at $w_0=-\arccosh(\bar r)$, yielding a gaussian decay,
\begin{eqnarray}\label{rhod}
\rho_D&=& A_D\,
\bar r^{-\frac{d+5}{2}} \left(1-\bar r^2\right)^{\frac{-(1+d)}{4}}
\exp(-R_{\rm \rm tf}^2 \lambda)\\
A_D&=& \pi^{-(d+1)/2} 2^{-(d+1/2)}{R_{\rm tf}^{-1}}\\
\lambda&=& \bar r \sqrt{\bar r^2-1}-\arccosh(\bar r).
\end{eqnarray}
Comparing the width of the saddle point with the distance between $w_0$ and the pole at $w=0$, one finds that this expression is valid when $\bar r\gsim1+R_{\rm \rm tf}^{-4/3}$.

The density profile in this region can also be estimated from a modified Gross-Pittaevskii type functional \cite{kolomeisky}.  However, such an approach can lead to unphysical interference  predictions \cite{unphys}.

\noindent {\bf Region C.} Between regions $B$ and $D$ ($1-1/R_{\rm tf}^{4/3}<\bar r^2<1+1/R_{\rm tf}^{4/3}$) all three saddle points are in close proximity to $t=1$ ($w=0$ in the variables used to discuss region D).  Linearizing $\bar \theta$ about $w=0$, one finds to leading order,
\begin{equation}\label{lint}
\bar\theta=-( r^2-R_{\rm tf}^2) w -(1+d/2)\log(w)- r^2 w^3/3.
\end{equation}
By introducing a small parameter, $\eta=(3^{1/3}  r^{-2/3} (R_{\rm tf}^2-r^2)),$ and changing integration variables to $v=r^2 w^3/3$, we can write the integrand in a power series in $\eta$.  Each term can be integrated by using a variant of equation (\ref{theorem}), yielding
\begin{eqnarray}\label{rhoc}
\rho_C &=&A_C r^{d/3}\sum_j \frac{\eta^j}{\Gamma(j+1)\Gamma(d/6+1-j/3)}\\
A_C&=&3^{-(1+d/6)}2^{-d} \pi^{-d/2}
\end{eqnarray}
This expansion is rapidly convergent when $\eta<1$, corresponding to 
$1-1/R_{\rm tf}^{4/3}<\bar r^2<1+1/R_{\rm tf}^{4/3}$.

%Figure~\ref{compare} illustrates the accuracy of this approximation, even for $M\sim 1$.  In Fig.~2 we compare our results to those of Brack and Murthy \cite{brack},
%\begin{equation}
%\delta\rho_{\rm BM} = \frac{(-1)^M}{2^{d}\pi^{d/2}}\bar r^{1-d/2} J_{d/2-1}(2\bar r R_{\rm \rm tf}^2),
%\end{equation}
%Where $J_{\mu}$ is the Bessel function of order $\mu$.
%For $d=1$ this also agrees with the result of Gleisberg et al \cite{glesberg}.  By expanding form small $\bar r$, one sees that these asymptotic expressions all agree for $x$ near $0$.  However, Eq.~(\ref{res1}) is clearly a more uniform approximation to the density oscillations.

\begin{figure}
\includegraphics[width=\columnwidth]{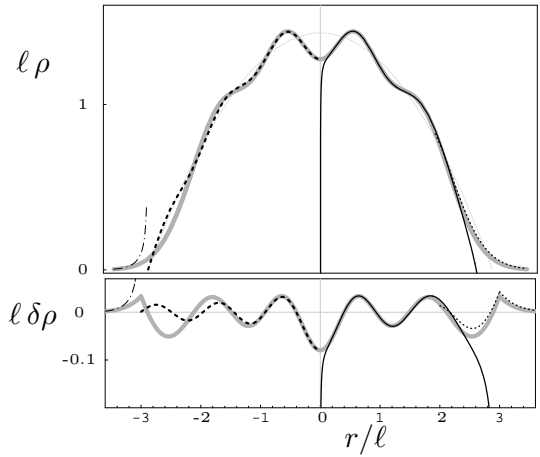}
\caption{
Density profile of four filled shells (M=3, N=10) of non-interacting fermions in a two dimensional harmonic trap.  Top and bottom panels respectively show density $\rho$ and deviation $\delta \rho$ from a Thomas-Fermi profile. 
For clarity, the various 
approximate solutions are shown only on the left or right half of the graph.
Thick gray line: exact; solid black line (right):  Eq. (\ref{res1}); dotted line (right): Eq. (\ref{rhoc}); dashed line (left): Eq. (\ref{rhoa}); dot-dashed line (left): Eq. (\ref{rhod}).
}\label{allap}
\end{figure}

\noindent {\bf Summary:}
We have analytically calculated the density profile for $N$ noninteracting spin-polarized fermions in a $d$-dimensional harmonic trap.  Four different expressions are needed to describe all parts of the cloud.  The results from all four regimes are all sketched in figure~\ref{allap} for the case of ten particles in two dimensions (corresponding to four filled shells).

%We find that the mean density scales as $N^{1/2}$ and that the leading order fluctuations scale as $N^{-1/2d}$.

\noindent {\bf Acknowledgments:}
This work was partially performed at the Kavli Institute for Theoretical Physics.  It was supported in part by the National Science Foundation under Grant No. PHY99-07949.


\begin{thebibliography}{99}
\bibitem{behavior}One requires that
the $n$'th derivative, $f^{(n)}(1)$, is of order $1$, and that $f(t)\to0$ sufficiently rapidly as $t\to\pm i \infty$.
\end{thebibliography}
  \end{document}